\documentclass[submission,copyright,creativecommons]{eptcs}
\providecommand{\event}{FROM 2026} 

\usepackage{iftex}

\ifpdf
\usepackage{underscore}         
\usepackage[T1]{fontenc}        
\else
\usepackage{breakurl}           
\fi

\title{Formalization of Fragments of the Theory \\of Hereditarily Finite Sets} 

\author{Zuzana Haniková
	\institute{Institute of Computer Science of the Czech Academy of Sciences}  \email{hanikova@cs.cas.cz}
	\and 
	Štěpán Holub
	\institute{Faculty of Mathematics and Physics, Charles University, Prague} \email{holub@karlin.mff.cuni.cz} 	   
	\thanks{Both authors acknowledge support of the Czech Science Foundation project no. 25-16489S}
}

\newcommand\titlerunning{Formalization of Hereditarily Finite Sets} 
\newcommand\authorrunning{Z. Haniková and Š. Holub}

\hypersetup{
	bookmarksnumbered,
	pdftitle    = {\titlerunning},
	pdfauthor   = {\authorrunning},
	pdfsubject  = {ZFfin},               
}

\usepackage[T1]{fontenc}
\usepackage{graphicx}

\usepackage{amsmath,amssymb}
\usepackage{enumitem}

\usepackage{xcolor}
\usepackage{isabelle-listings}

\usepackage{tikz}
\usetikzlibrary{arrows.meta,bending,shapes.misc}

\lstnewenvironment{Icode}[1][] 
{\lstset{#1}}                
{}
\newcommand{\axitem}[1]{\item[{\tt (#1)}]}

\newcommand{\theory}[1]{\ensuremath{\mathrm{#1}}}
\newcommand{\atm}{\theory{AST}}
\newcommand{\zf}{\theory{ZF}}
\newcommand{\zfc}{\theory{ZFC}}
\newcommand{\zffin}{\theory{{ZF}_{\mathrm{fin}}}}

\newcommand{\pa}{\theory{PA}}
\newcommand{\qa}{\theory{Q}}

\newcommand{\THF}{the theory of hereditarily finite sets}
\newcommand{\SFP}{\texttt{SetFormulaPredicate}}

\long\def\snote#1{\bgroup\color{blue} (#1) \egroup}
\long\def\znote#1{\bgroup\color{magenta} (#1) \egroup}

\definecolor{atomcol}{RGB}{210,228,246}    
\definecolor{compcol}{RGB}{252,244,214}    
\definecolor{sublocalecol}{RGB}{40,90,170}      
\definecolor{indcol}{RGB}{170,30,30}      

\tikzset{
    x=1in,y=1in,
	every node/.style={font=\ttfamily\scriptsize},
	lnode/.style={draw,rounded corners=1.4pt,minimum height=15pt,line width=0.3pt,fill=white,anchor=center},
	atomicN/.style={lnode,fill=atomcol},
	compoundN/.style={lnode,fill=compcol},
	sublocaleE/.style={-{Latex[length=4pt]},shorten >=3pt,line width=0.6pt,sublocalecol},
	equivE/.style={{Latex[length=5pt]}-{Latex[length=5pt]},shorten >=3pt,shorten <=3pt,line width=0.6pt,sublocalecol},
    indE/.style={-{Latex[length=4pt]},shorten >=3pt,line width=0.45pt,indcol},
	indN/.style={midway, ultra thick, fill = white, inner sep = 0pt}
}

\begin{document}
	\maketitle

\begin{abstract}
The axiomatization of the theory of hereditarily finite sets in first-order classical logic is systematically explored and formalized in Isabelle/HOL. The formalization uses a hierarchy of locales, each corresponding to a fragment of the theory given by a particular collection of axioms.
An inductive definition of first-order definable predicates is introduced and used to formalize axiom schemata. Special attention is paid to several equivalent axioms of finiteness, as well as to several equivalent ways of expressing regularity. The work also formalizes several facts about independence of an axiom from a system of axioms by defining appropriate models. 
\end{abstract}

\section{Introduction}
\label{s:intro}
The theory of hereditarily finite sets currently lacks a comprehensive exposition, despite numerous papers exploring it and its fragments. To fill this gap, we use the proof assistant Isabelle/HOL to systematize the hierarchy of its fragments. The approach is modular, admitting later additions and refinements, and is also relevant in the case of $\zf$ (if the axiom of infinity is assumed). The formalization uses a hierarchy of locales, each corresponding to a fragment of the theory given by a particular collection of axioms.

Without the axiom of infinity, it may come as a surprise that the usual regularity axiom in itself does not guarantee well-foundedness of the membership relation, that is, the absence of infinite descending chains. 
In particular, it is not a given that each set has a transitive closure (notice that a descending chain is the transitive closure of its first element, and if the chain is infinite, so is the closure). This raises one of the intriguing questions, formally explored in this paper, of axiomatically guaranteeing well-foundedness of hereditarily finite sets. The hereditarily finite setting also yields an interesting interplay between the axiom schemata of replacement and of separation; they appear comparable in strength, and in some contexts, separation entails replacement. Another point addressed is the different ways of defining finiteness (see Section \ref{ss:neginf}). 
Their interplay is exposed in weaker contexts, when only fragments of the theory are available. The theory of hereditarily finite sets also highlights the importance of (different versions of) the axiom schema of induction, which is, in fact, a prominent way of defining finiteness.

The interest in the theory of hereditarily finite sets also follows from its being bi-interpretable with Peano arithmetic. As a side note, a first inspiration for our work, which is nevertheless not pursued in this paper, comes from Vopěnka's Alternative Set Theory ($\atm$) as presented in
\cite{Vopenka:Teubner} and in a sizeable series of papers including \cite{Pudlak-Sochor:ModelsAST,Sochor:BasesAST,Sochor-Vopenka:Shiftings,Sochor:AST76,Sochor:ASTmeta-I,Sochor:ASTmeta-II,Sochor:ASTmeta-III}. The $\atm$ is a theory of sets and classes, in some aspects analogous to Gödel--Bernays or Kelley--Morse set theories (where classes are not shortcuts, as they are in $\zf$). Vopěnka's $\atm$ is related to our project as it presents a radically new theory of infinity, which is informally motivated by the notion of clarity. All sets are formally finite, but they can contain subclasses that are themselves not sets (called \emph{proper semisets}), which---as Vopěnka claims---capture indefiniteness. 

The code is published as \cite{HolubHanikova:AxiomaticZFfinite}.
The state of the code, compatible with Isabelle-2025-2 release, to which we refer in this paper will be permanently available in a public repository \cite{Zffinrepo}. 

Given that we deal with a classical topic, all the main results we present are covered by existing literature, albeit scattered across different sources,
some of which are difficult to find. It is not generally known that the literature about $\atm$ subsumes a development of \THF.
While some familiarity with the $\atm$ literature will be apparent from our list of references, we try to reference the first published record of each formalized fact to our best knowledge. 
Unifying the varied resources and bringing them under a common roof has required, for example, a careful examination of the exact formulation of the axioms. We provide a systematic and verified presentation, which in particular allows for a  reliable account of coincidences or distinctions of the various fragments. In other words, we obtain and represent information about the ordering of those fragments of \THF{} by their strength. Let us illustrate this by an example.
A result concerning dependence of two axioms in a particular context are the following two theorems (see 
Section \ref{s:axioms} for the meaning of the acronyms, Subsection \ref{ss:reg} for detailed exposition of mathematical facts, and ‹Summary of dependencies› in \texttt{ZFfin.thy} for the formalization):
\begin{Icode}
 theorem  (in L_setext_sep_reg) ts_implies_epsind: 
  "L_ts (\<epsilon>) \<Longrightarrow> L_epsind (\<epsilon>)"	
 theorem  (in L_setext_empty_union_repl_pair) epsind_implies_ts:
  "L_epsind (\<epsilon>) \<Longrightarrow> L_ts (\<epsilon>)"
\end{Icode}
which are related to \cite[Proposition 5.4.]{Kaye-Wong:IntpArithmeticSetTheory}.  The cited proposition reads\footnote{Elementary Set Theory, introduced in \cite{Baratella-Ferro:theoryNegInf},
	is axiomatized by the axioms of extensionality, empty set, pairing, union, and the replacement schema. Found denotes the axiom of regularity in \cite{Kaye-Wong:IntpArithmeticSetTheory}.}: 
	\medskip  

\textbf{Proposition 5.4}\ \textit{For all} $V \models \text{EST} + \text{Found}$,
\[  V \models \text{$\in$-Ind} \Leftrightarrow V \models \text{TC}.\]

\noindent Our theorems prove each of the two implications separately in an appropriate context, which seems a natural way to go when formalizing the fact, since
eventually the equivalence is obtained explicitly (almost) for free in a sufficiently strong  context just by exploring the locale hierarchy.

The project is also interesting from the formalization point of view, as an attempt to capture a first-order theory and its properties (including its models) in Isabelle/HOL. A particular challenge stems from the fact that \THF{} is not finitely axiomatizable (neither is Peano arithmetic for that matter), so it is necessary to address the problem of restricting the predicates that can be used in axiom schemata to those that are definable by first-order formulas in the given formal language (in our case, the language with membership and equality).
We present a solution to that challenge which can be called \emph{semantic}. We do not formalize formulas as such, 
but rather the properties they define, introducing an inductive definition of those properties that are set-theoretically definable.

The formalization uses Isabelle/HOL locales to represent fragments of set theory given by particular axioms.
The locales are named in a recognizable way using the names of the axioms or schemata whose combination they represent.
There are two exceptions: the locale {\tt ZFfin} represents the theory whose intended model is $V_\omega$, obtained by
modifying the theory $\zf$, and the locale {\tt ASTset}, which represents
the set fragment of the theory $\atm$. We eventually show equivalence of these two distinct ways of axiomatizing \THF{}.

Our formalization includes two distinct kinds of mathematical facts. Firstly, we formalize ``positive facts''
about the provability of statements in various fragments. These facts allow us to refine the hierarchy of locale dependencies.

Secondly, we formalize several ``negative facts'', stating the independence of a statement from a certain collection of statements. The latter results are a different kind of challenge, given that they require constructing appropriate (counter-)models. One way to obtain such models is by defining a new membership relation using a definable permutation on sets. We formalize this Rieger--Bernays permutation approach (see \cite{Forster:PermutationModelsRB} for an exposition and attribution), and apply it to show independence of the axiom of regularity and the schema of regularity from certain other axioms.
For example, we formalize a construction of a model which exemplifies the situation mentioned above, namely satisfying the standard axiom of regularity while not being well-founded.

Three figures survey positive facts included in the formalization.  
Arrows indicate dependencies between locales formalized with the \texttt{sublocale} command. We display only dependencies that require a nontrivial proof. We omit facts that follow from transitivity and locale extensions, that is, those discharged by mere \texttt{unfold\_locales} (see the end of Section \ref{ss:neginf} for an example). Independence results are summarized in Figure \ref{fig:graph-independence} using crossed-out arrows.

As it stands, the hierarchy of locales spans more than 3000 lines of code.
It is not the intent of this work to exhaust the space of facts that could in principle be obtained.
Rather, we focus on designing a formalization approach that is convenient
for such an endeavour and then on formalizing the equivalence of several finiteness principles, and of several ways of expressing regularity, each in appropriate axiomatic context.

%

This paper is organized as follows.
Section \ref{s:prelim} collects a modicum of definitions from logic and set theory.
Section \ref{s:axioms}  provides an overview of the precise formulations of axioms we use.
Section \ref{s:high}  presents the main mathematical facts covered by the formalization.
Section \ref{s:formal} explains the design ideas of the formalization.
Section \ref{s:related} discusses relevant previous formalization work.
The last Section \ref{s:final} outlines possible future directions.

\section{Logic and set theory preliminaries}
\label{s:prelim}

Theories formalized in this work are first-order (one-sorted) set theories in classical logic with equality.
In particular, the equality symbol $=$ is always available and counts as a logical symbol.
The basic language of the theories has only one symbol $\in$ (we use $\varepsilon$ in the formalization to distinguish the abstract relation from the membership of Isabelle/HOL, and this is reflected in the reproduced code; but we stick with the more common $\in$ for the presentation here, thus $\in$ and $\varepsilon$ need to be viewed as interchangeable within the text of this paper). 
Congruence axioms for $\in$ w.r.t.~$=$ are included among logical axioms. 
Any other function or predicate symbols that occur are definable by means of $\in$ and $=$.

A set formula is any formula in the above language (possibly featuring some defined symbols).
If $M$ is a structure for the language of set theory and $R$ is a relation in $M$,
we say that $R$ is set-theoretically definable provided there is a set formula $\varphi$ that defines $R$ in $M$. 

When developing an axiomatic theory of sets in first-order one-sorted logic, 
it might seem superfluous to talk about \emph{set} formulas: 
 under the given setup, any formula is a set formula by necessity. 
On the other hand, one can write down various expressions, and define various relations, in Isabelle/HOL,
so it will be our prime concern to indicate those that represent set formulas and set-theoretically definable relations.
 
An important example of a situation requiring such a distinction is axiom schemata in $\zf$-like set theories (or any non-finitely axiomatizable theories that use axiom schemata in some fixed  language), such as separation, replacement, 
induction, and $\in$-induction. 

Another reason to take care about the set language in particular is a possible further expansion of the theory of sets
 to a theory of sets and classes (namely the $\atm$), which could be conceived as a two-sorted theory in a similar way in which second-order arithmetic
 can be so conceived.

Zermelo-Fraenkel set theory $\zf$ has the following axioms and schemata: 
extensionality, empty set, union, powerset, axiom schema of replacement, infinity, and regularity.
The axiom of pairing and the axiom schema of separation are derivable in $\zf$.
We use $\bigcup$ for (unary) union operation and $\mathfrak{P}$ for powerset operation. 

The theory $\zffin$ is axiomatized by extensionality, empty set, union, powerset, axiom schema of replacement, 
axiom schema of $\in$-induction, and the negation of infinity. 

A map in $\zf$ is a set or class of ordered pairs (encoded in the usual way). For any two sets $x$ and $y$, a set bijection 
of $x$ onto $y$ is an injective set map $f$ (i.e., $f$ is a set) of $x$ onto $y$;
if such a map exists, we write $x\mathrel{\hat\approx} y$. 
A set $x$ is Dedekind finite if and only if there is no proper subset $y\subset x$ such that $y\mathrel{\hat\approx} x$.
A set $x$ is Tarski finite if and only if any nonempty subset of $\mathfrak{P}(x)$ has a maximal element under inclusion.

A set $x$ is transitive (write $\mathrm{trans}(x)$) if and only if $y\in x$ entails $y\subseteq x$. 
A set $x$ is an ordinal number if and only if it is transitive and totally well-ordered by $\in$
(that is, $\forall y\,z \in x (y\in z \lor y=z \lor z\in y)$ and any nonempty subset of $x$ has a $\in$-least element).  
A natural number is an ordinal $x$ such that $x$ and each of its elements are either empty, or they are successor ordinals.

In the formalization, we use the subscript \texttt M uniformly as a means of avoiding a conflict with existing notation. 
We therefore have $\bigcup_{\texttt M}$, $\mathfrak{P_{\texttt{M}}}$, $\subseteq_{\texttt M}$, etc. Using $\varepsilon$ instead of $\in_{\texttt M}$ is an exception.

\section{List of axioms used}
\label{s:axioms}

Some but not quite all axioms used in the formalization are well known. 
Below is a complete list, providing for each of its items a full name and an abbreviated one, where the latter is used as the name of the corresponding locale in the formalization. 
Some axioms are presented in a language expanded with new symbols (such as $\emptyset$, $\{y\}$, $x \cup y$ or $x \mathrel{\hat\approx} y$). In appropriate contexts these symbols become definable, and can be eliminated. (See the discussion on Hilbert's definite description operator available in HOL, in Section \ref{s:formal}.) 

The convention for free variables is to list them explicitly.\footnote{With any set formula $\alpha$, writing $\alpha(x_1,\dots,x_n)$ for a list of variables $x_1,\dots, x_n$ for some natural number $n$ means that the list  $x_1,\dots, x_n$ subsumes all free variables  in $\alpha$. In particular $\alpha(x_1,\dots,x_n)$ need not have any free variables.} 
A list of variables $x_1,\dots, x_n$ can be written as $\bar x$.   
 The formula $\varphi(x_m,x_2,\dots,x_n)$ for $m>n$ is a substitution instance of a formula  $\varphi(x_1,\dots,x_n)$  obtained by uniformly replacing each occurrence of $x_1$ with $x_m$. 

\begin{itemize}[leftmargin=2.6cm]
	\axitem{setext} axiom of extensionality for sets (A11 in \cite{Sochor:ASTmeta-I})
	$$\forall x\,y\, (x=y \leftrightarrow \forall u (u\in x \leftrightarrow u\in y))$$ 
	\axitem{empty} axiom of empty set
	$$\exists z\,\forall  u\, \neg (u\in z)$$
	\axitem{setsuc} axiom of set successor 
	$$ \forall x\,y\, \exists z\, \forall u\, (u\in z \leftrightarrow u\in x \lor u=y) $$
	\axitem{pair} axiom of pairing 
	$$ \forall x\,y\, \exists z\, \forall u\, (u\in z \leftrightarrow u\in x \lor u\in y) $$
	\axitem{setind} axiom schema of induction for set formulas.\footnote{Cf.~\cite[p.~19]{Vopenka:Teubner} (Vopěnka omits the free variables in the statement, but assumes their presence in various instantiations of the schema). See also A41 in \cite{Sochor:ASTmeta-I}}      
	For any set formula $\varphi(x,\bar w)$, 
	$$ \forall \bar w\, (\varphi(\emptyset,\bar w) \land \forall x\,y\,(\varphi(x,\bar w)\to\varphi(x\cup\{y\},\bar w)) 
	\to \forall x\,\varphi(x,\bar w)) $$ is an axiom.
	\axitem{regsch} axiom schema of regularity for set formulas. 
    For any set formula $\varphi(x,\bar w)$,
	$$ \forall \bar w\, ( \exists x\, \varphi(x,\bar w) \to \exists x\, (\varphi(x,\bar w) \land \forall y\, (y\in x \to \neg\varphi(y,\bar w))))$$
	 is an axiom.
	\axitem{epsind} axiom schema of $\in$-induction for set formulas. 
    For any formula $\varphi(x,\bar w)$,
	$$ \forall \bar w\,  (  \forall x\, (\forall y\, ( y\in x \to \varphi(y,\bar w)) \to \varphi(x,\bar w) ) \to \forall x\, \varphi(x,\bar w) ) $$
	is an axiom.
	\axitem{reg} regularity 
	$$\forall x\, ( \exists z\, (z\in x) \to \exists z\, (z \in x \land \forall u\, (u\in z \to \neg (u\in x)) ) $$
	\axitem{sep} axiom schema of separation for set formulas. For any set formula  $\varphi(u,\bar w)$, 
	$$ \forall \bar w\, ( \forall x\, \exists z\, \forall u\, (u \in z \leftrightarrow  u\in x \land \varphi(u,\bar w))) $$ 
	is an axiom.
	\axitem{repl} axiom schema of replacement for set formulas. For any set formula  $\varphi(u,v,\bar w)$, 
	$$\forall \bar w\, [  \forall u\, \exists ! v\, \varphi(u,v,\bar w) \to 
		\forall x\, \exists z\, \forall v\, (v\in z \leftrightarrow \exists u\, (u\in x \land \varphi(u,v,\bar w)) ) ] $$  
    \axitem{union} axiom of union 
	$$ \forall x\, \exists z\, \forall u\, ( u\in z \leftrightarrow \exists y\,  ( u\in y \land y\in x ) ) $$    
	\axitem{power} axiom of powerset
	$$ \forall x\, \exists z\, \forall u\, (u\in z \leftrightarrow u\subseteq x)$$
	\axitem{inf} axiom of infinity 
	$$\exists x\, (\emptyset \in x \land \forall y \in x (y\cup\{y\} \in x ) )$$
        \axitem{fin} axiom of finiteness $=\neg\mbox{{\tt (inf)}}$	
        \axitem{tarski} axiom of Tarski finiteness 
	 $$ \forall x\, y\, [(y\not=\emptyset \land  \forall z\,  (  z\in y \to z\subseteq x ) ) \to 	\exists z\, (z\in y \land \neg (\exists w\, (w \in y \land z \subset w))) ]
 	$$
 	\axitem{dedekind} axiom of Dedekind finiteness
        $$ \forall x\, y\, (y\subset x \to \neg (x \mathrel{\hat\approx} y)) $$
        \axitem{ts} axiom of transitive superset 
        $$ \forall x\, \exists z\, (\mathrm{trans}(z) \land x \subseteq z)$$
        \axitem{setindregsch}  axiom schema of induction and regularity for set formulas, cf.~\cite{Pudlak-Sochor:ModelsAST}.
         For any set formula $\varphi(x,\bar w)$, 
	$$ \forall \bar w\, (\varphi(\emptyset,\bar w) \land \forall x\,y\,(\varphi(x,\bar w)\land (\varphi(y,\bar w) \to\varphi(x\cup\{y\},\bar w)) 
	\to \forall x\,\varphi(x,\bar w)) $$ is an axiom. 
        \end{itemize}

\section{Main formalized results}
\label{s:high}

This section summarizes interesting results included in the
formalization, along with pointers to literature where these results appeared in print.
We also point out a few omissions and redundancies.

A fair amount of technical development takes place in fragments that each provide a natural, sufficiently strong milieu for the required purpose, but at the same time does not stand in the limelight. Several such fragments are worth mentioning here. The first one is axiomatized by extensionality, empty set, and set successor. This is a rather weak theory (subsumed by all the other fragments mentioned in this paragraph) which nevertheless plays an important role in logic. For example, it was shown by Tarski and Szmielew to interpret Robinson's arithmetic $\qa$ (see \cite{Tarski-Mostowski-Robinson:Undecidable}; in fact this fragment is mutually interpretable with Q). The second fragment is axiomatized by extensionality, empty set, powerset, and replacement.
This theory is rich enough to prove useful lemmas about functions, bijections, and cardinality, as well as equivalence of all finiteness principles we will consider below. Yet another is Elementary Set Theory (EST) as in \cite{Baratella-Ferro:theoryNegInf},
axiomatized by extensionality, empty set, pairing, union, and the replacement schema. Lastly, there is the fragment axiomatized by extensionality, empty set, set successor, and the set induction schema; if the schema of regularity is added on top of it, one obtains the set fragment of the theory $\atm$, as in \cite[Chapter I, Section 1]{Vopenka:Teubner}.

\subsection{Negating the axiom of infinity}
\label{ss:neginf}

In $\zf$ and in  $\zfc$, there are various ways to express the fact that a set is finite (see e.g.~\cite{Levy:noteDefFiniteness}).
We are interested in a different albeit related concept, namely
for \emph{each} set to be finite (and hence, hereditarily finite).
Many different principles can be found that convey this design choice axiomatically.
To be more precise, any such principle will guarantee finiteness of each set in the universe
when it is considered in a suitable context---that is, in the presence of a few other axioms.
The context may differ for different finiteness principles.
We consider four principles: axiom schema of induction, negation of the axiom of infinity, Tarski finiteness, and Dedekind finiteness.

\smallbreak\noindent
{\bf A.~Axiom schema of induction for set formulas} \texttt{(setind)}.
We will start from axiom schema of induction for set formulas considered as a finiteness principle.
It is shown in \cite[Chapter I, Section 1]{Vopenka:Teubner} that a theory with 
extensionality {\tt (setext)}, empty set {\tt (empty)}, set successor {\tt (setsuc)}, and axiom schema of induction for set formulas 
{\tt (setind)} proves 
the axioms of union, powerset, and the axiom schema of replacement, as well as
the pairing axiom and the schema of separation (derivable from the former). 
Further, each set has a minimal and max\-imal element under inclusion, 
in particular any nonempty subset of the powerset of any given set $x$ has an inclusion-maximal element and  $x$ is Tarski finite.
It is then easy to prove the negation of the axiom of infinity by contraposition ({\tt sublocale\ L\_fin}), since 
an inductive set is not Tarski finite. Induction also entails that every set is Dedekind finite. 
All these facts are available in \cite{Vopenka:Teubner}. 
The locale  {\tt L\_setext\_empty\_setsuc\_setind} mirrors this axiomatization, and one can prove each of
the corresponding sublocale commands (such as ${\tt sublocale\ L\_power}$, ${\tt sublocale\ L\_repl}$, etc.) therein. 

Thus we see that induction for set formulas, considered in the context of extensionality, empty set and set successor axioms, is a powerful statement, responsible for a lot of heavy lifting. It proves several of the axioms of $\zf$, and it can be taken as a finiteness principle:
it proves the other principles {\bf B}, {\bf C}, and {\bf D} listed below, 
and it also proves their appropriate contexts as required.

Notice that none of the claims above require any regularity axioms.

\smallbreak\noindent
{\bf B.~Negated axiom of infinity} \texttt{(fin)}.
A natural approach to axiomatizing hereditarily finite sets is removing the axiom of infinity {\tt (inf)} from 
the system of axioms of $\zf$
and replacing it with its negation {\tt (fin)}.
It is well known, to the point of being folklore (cf.~\cite{Vopenka-Hajek:Fundierungsaxiom, Kaye-Wong:IntpArithmeticSetTheory}) that not all sets necessarily have a transitive superset in this theory (i.e., flipping {\tt (inf)} into {\tt (fin)}). 
The problem can be addressed by using the schema {\tt (epsind)} instead of {\tt (reg)}, as discussed in Subsection \ref{ss:reg} below.
This yields the theory we denote $\zffin$, following in the footsteps of \cite{Sochor:ASTmeta-I, Sochor:ASTmeta-II,Sochor:ASTmeta-III}. The axiomatization is mirrored by the locale {\tt ZFfin} in subsection ‹Summary of dependencies› of \texttt{ZFfin.thy},
but many interesting results can be proved from weaker fragments of this theory already.

To demonstrate other finiteness principles in the theory with {\tt (fin)} as a finiteness principle, 
it is convenient to first prove schema of induction for set formulas in it. 

\begin{figure}
	\centering
	\begin{tikzpicture}
		\node[atomicN] (Lempty) at (1,4.5) {L\_empty};
		\node[atomicN] (Lsep) at (1,4) {L\_sep};
		\node[compoundN] (Lsetextemptyrepl) at (1,3.5) {L\_setext\_empty\_repl};
		\node[atomicN] (Lbinunion) at (2.3,4.5) {L\_binunion};
		\node[compoundN] (Lsetextemptyunionreplpair) at (2.3,4) {L\_setext\_empty\_union\_repl\_pair};
		\node[atomicN] (Lpair) at (3.35,4.5)  {L\_pair};
		\node[compoundN] (Lemptysetsuc) at (3.35,3.5) {L\_empty\_setsuc};
		\node[atomicN] (Lsetsuc) at (5,4.5) {L\_setsuc};
		\node[compoundN] (Lsetextpairbinunion) at (4.15,4) {L\_setext\_pair\_binunion};
		\node[compoundN] (Lsetextemptypowerrepl) at (5.75,4) {L\_setext\_empty\_power\_repl};
		\draw[sublocaleE] (Lpair) -- (Lemptysetsuc);
		\draw[sublocaleE] (Lempty) -- (Lsep);
		\draw[sublocaleE] (Lsep) -- (Lsetextemptyrepl);
		\draw[sublocaleE] (Lsetsuc) -- (Lsetextemptypowerrepl);
		\draw[sublocaleE] (Lsetsuc) -- (Lsetextpairbinunion);
		\draw[sublocaleE] (Lbinunion) -- (Lsetextemptyunionreplpair);
	\end{tikzpicture}
	\caption{Additional proven dependencies between basic axioms.
	}
	\label{fig:graph-basic}
\end{figure}
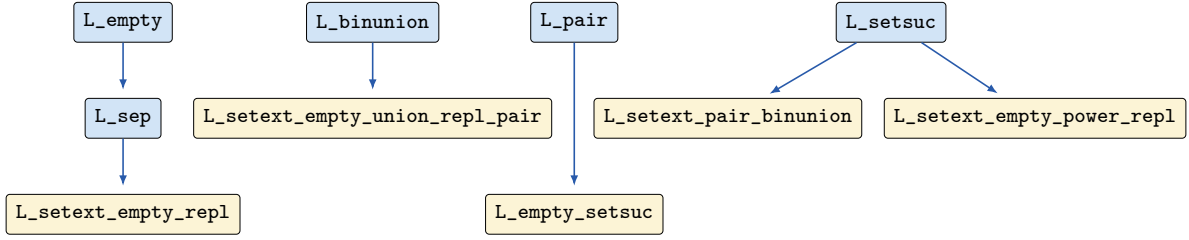

To demonstrate other finiteness principles in the theory with {\tt (fin)} as a finiteness principle, 
it is convenient to first prove schema of induction for set formulas in it. 
Our proof builds on a portion of the theory  of natural numbers developed in the fragment axiomatized by extensionality, empty set, powerset, and replacement mentioned at the beginning of this section (subsection ‹Powerset› of \texttt{ZFfin.thy}). 
Assuming, in addition, our axiom of interest, {\tt (fin)},   
 it can be established that the cardinality relation is a function, i.e., for each set $x$, there is at most one 
natural number such that  a set bijection exists between $x$ and $n$.  
Now $\emptyset$ has cardinality and for any sets $a$ and $b$, if $a$ has cardinality $n$, then $a\cup\{b\}$ has cardinality $n+1$ provided $b\not\in a$, so having cardinality is closed under taking set successors.
Suppose for an arbitrary property $P$ defined by a set formula that a set $x$ violates the instance of induction for $P$: then 
take those elements of $\mathfrak{P}(x)$ that satisfy $P$, and using replacement on the relation of having cardinality, 
obtain an inductive set containing $\emptyset$, which contradicts {\tt (fin)}.
In this way, we prove {\tt sublocale L\_setind} and establish {\bf A} in the locale {\tt L\_setext\_empty\_power\_repl\_fin} (see the subsection ‹Negation of inf› of \texttt{ZFfin.thy}).
As a consequence, the axiom of union is provable, since \texttt{(setsuc)} is provable  (see Figure \ref{fig:graph-basic}), and, as discussed above, \texttt{(union)} is provable with the use of induction for set formulas in the context of extensionality, empty set, and set successor.


\smallbreak\noindent
{\bf C. Tarski finiteness} \texttt{(tarski)}.
Now the finiteness assumption is that each set is Tarski finite. 
Again we would like to prove the induction schema (that is, to establish {\bf A}),
so the question is which additional axioms to assume as a context.
Our proof follows \cite[p.~60]{Sochor:ASTmeta-II}, even if its author claims to be working from the assumption 
{\tt (fin)} rather than {\tt (tarski)}.  

As in {\bf B} above, assuming an inductive predicate $P$ and a set $x$ not satisfying $P$, we obtain a subset $z$ of $\mathfrak{P}(x)$ that violates the Tarski finiteness, namely the set of subsets of $x$ that satisfy $P$. Since $\neg P(x)$, any element of $z$ can be made larger, and $z$ has no maximal element under inclusion. No replacement is needed in this proof,  
the context is extensionality, empty set, successor, separation, and powerset.
(Formally, we work in the locale  {\tt L\_setext\_empty\_power\_sep\_setsuc\_tarski}.)  
Thus we have established {\bf A}.

As above, the provability of induction in this axiomatization entails that replacement {\tt (repl)} and {\tt (union)} are provable too.
This provides a perspective from which the separation schema is at least as strong as the replacement schema.
These provabilities have been shown explicitly  in \cite{Behounek:nezavislostAx}, cf.~2.1 (replacement) and 4.3 (union).
Furthermore, that work presents a model (not represented within the current state of the formalization) which showcases 
that if  the axiom {\tt (fin)} is assumed instead of {\tt (tarski)}, then even with union, replacement cannot be derived from separation.
These facts can be taken as finer points in assessing  the behaviour of the separation schema and the replacment schema in 
theories that subsume some finiteness principle (here, Tarski finiteness and negation of infinity).

\smallbreak\noindent
{\bf D. Dedekind finiteness} \texttt{(dedekind)}.
Our exposition again starts from the fragment axiomatized by extensionality, empty set, powerset, and replacement.
If one furthermore assumes {\tt (dedekind)}, the axiom stating that each set is dedekind finite, then the axiom ${\tt (fin)}$ is obtained as a theorem.
This can be established by contradiction: assuming  the existence of an inductive set, 
one can separate the set $\omega$ of natural numbers. Then a  set bijection of $\omega$ to $\omega\setminus\{\emptyset\}$, 
sending  each natural number to its successor, is obtained by separation from $\omega^2$,
which shows $\omega$ to be Dedekind infinite. 
Hence in the given context we prove the axiom {\tt (fin)} of {\bf B}. 

\smallbreak
Altogether, we have that in the context of \texttt{(setext)}, \texttt{(empty)}, \texttt{(power)} and \texttt{(repl)}, all four finiteness principles are equivalent. 

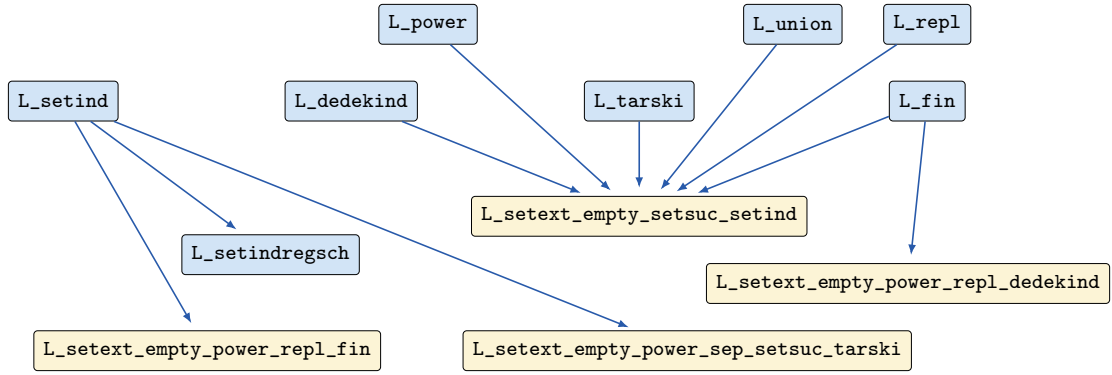
\begin{figure}
	\centering
\begin{tikzpicture}
	\node[atomicN] (Lsetind) at (1.5,3.1) {L\_setind};
	\node[atomicN] (Ldedekind) at (3,3.1) {L\_dedekind};
	\node[atomicN] (Ltarski) at (4.5,3.1) {L\_tarski};
	\node[atomicN] (Lfin) at (6,3.1) {L\_fin};
	\node[atomicN] (Lrepl) at (6,3.5) {L\_repl};
	\node[atomicN] (Lunion) at (5.3,3.5) {L\_union}; 
	\node[atomicN] (Lpower) at (3.4,3.5) {L\_power};
	\node[compoundN] (Lsetextemptysetsucsetind) at (4.5,2.5) {L\_setext\_empty\_setsuc\_setind};
	\node[compoundN] (Lsetextemptypowerunionrepldedekind) at (5.9,2.15) {L\_setext\_empty\_power\_repl\_dedekind};
	\node[atomicN] (Lsetindregsch) at (2.58,2.3) {L\_setindregsch};
	\node[compoundN] (Lsetextemptypowersepsetsuctarski) at (4.75,1.8) {L\_setext\_empty\_power\_sep\_setsuc\_tarski};
	\node[compoundN] (Lsetextemptypowerunionreplregfin) at (2.25,1.8) {L\_setext\_empty\_power\_repl\_fin};
	\draw[sublocaleE] (Ldedekind) -- (Lsetextemptysetsucsetind);
	\draw[sublocaleE] (Lpower) -- (Lsetextemptysetsucsetind);
	\draw[sublocaleE] (Lfin) -- (Lsetextemptypowerunionrepldedekind);
	\draw[sublocaleE] (Lfin) -- (Lsetextemptysetsucsetind);
	\draw[sublocaleE] (Lrepl) -- (Lsetextemptysetsucsetind);
	\draw[sublocaleE] (Lsetind) -- (Lsetextemptypowersepsetsuctarski);
	\draw[sublocaleE] (Lsetind) -- (Lsetextemptypowerunionreplregfin);
	\draw[sublocaleE] (Lsetind) -- (Lsetindregsch);
	\draw[sublocaleE] (Ltarski) -- (Lsetextemptysetsucsetind);
	\draw[sublocaleE] (Lunion) -- (Lsetextemptysetsucsetind);
\end{tikzpicture}
	\caption{Proven dependencies between finiteness locales.
	}
	\label{fig:graph-fin}
\end{figure}

Formalized results about finiteness principles are summarized by Figure \ref{fig:graph-fin}.  For example, the fact that \texttt{(dedekind)} proves \texttt{(tarski)} in the context mentioned in the previous paragraph can be derived by a combination of displayed results as follows: \texttt{L\_setext\_empty\_power\_repl\_dedekind} proves \texttt{L\_fin},  \texttt{L\_setext\_empty\_power\_repl\_fin} proves \texttt{L\_setind}, and \texttt{L\_setext\_empty\_setsuc\_setind} proves \texttt{L\_tarski}, where \texttt{L\_setsuc} follows from \texttt{L\_setext\_empty\_power\_repl} (Figure \ref{fig:graph-basic}). Such ``obvious'' derivations correspond to proofs \texttt{by unfold\_locales} in the formalization. 
%

\subsection{Regularity and related principles}
\label{ss:reg}
 
The theories $\zf$ and $\zfc$ commonly use the axiom of regularity {\tt (reg)} to express the fact that membership in each model is a well-founded relation. In the context of all the other axioms of $\zf$, the same effect is achieved by using either the schema of regularity {\tt (regsch)} or the schema of $\in$-induction {\tt (epsind)} instead of regularity.
Notice that in a setting where one can talk about classes (denoted by uppercase variables), the schema of regularity can naturally take the form
\[ 
\exists x\, (x\in X) \to\ \exists x\, (x\in X\, \land\, x\cap X=\emptyset)\,.
\]



\begin{figure} 
	\centering
\begin{tikzpicture}
	\node[compoundN] (Lsetextemptysetsucsetindregsch) at (0.7,1.5) {L\_setext\_empty\_setsuc\_setindregsch};
	\node[atomicN] (Lreg) at (2,2.5) {L\_reg};
	\node[atomicN] (Lregsch) at (2,2) {L\_regsch};
	\node[atomicN] (Lepsind) at (1,2) {L\_epsind};
	\node[compoundN] (Lsetextsepregts) at (2.5,1.5) {L\_setext\_sep\_reg\_ts};
	\node[atomicN] (Lts) at (3.8,2.3) {L\_ts};
	\node[compoundN] (Lsetextemptyunionreplpairregsch) at (3.8,1.85) {L\_setext\_empty\_union\_repl\_pair\_regsch};
	\node[atomicN] (Lsetindregsch) at (3.8,1.3) {L\_setindregsch};
	\node[compoundN] (Lsetextsetsucsetindepsind) at (3.8,.8) {L\_setext\_setsuc\_setind\_epsind};  
	\draw[sublocaleE] (Lsetindregsch) -- (Lsetextsetsucsetindepsind);
	\draw[sublocaleE] (Lts) -- (Lsetextemptyunionreplpairregsch);
	\draw[equivE] (Lepsind) -- (Lregsch);
	\draw[sublocaleE] (Lepsind) -- (Lsetextemptysetsucsetindregsch);
	\draw[sublocaleE] (Lreg) -- (Lregsch);
	\draw[sublocaleE] (Lregsch) -- (Lsetextsepregts);
\end{tikzpicture}
	\caption{Proven dependencies between regularity and induction axioms.}	
	\label{fig:graph-reg}
\end{figure}
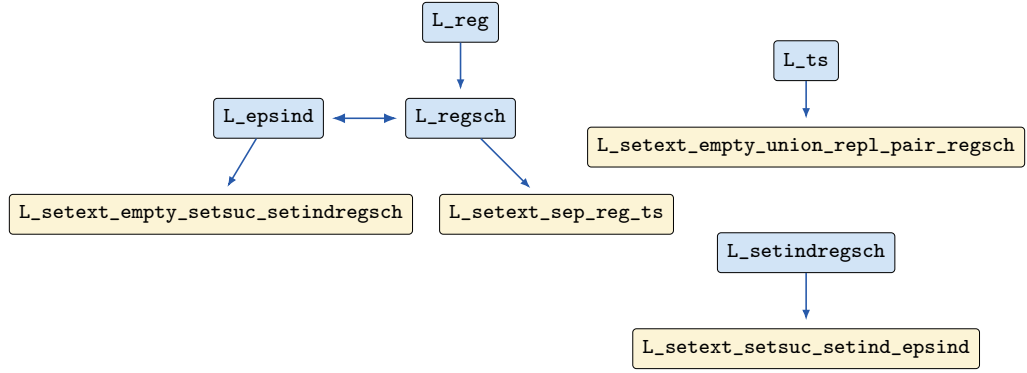

Our formalization explores four interrelated principles. For the independence claims below, see also subsection \ref{s:perm}.

\smallbreak\noindent
{\bf 1. Axiom of regularity} {\tt (reg)}, labelled A81 in \cite{Sochor:ASTmeta-I}, is independent of the fragment axiomatized by 
extensionality, empty set, set successor, induction schema for set formulas, and the axiom of transitive superset;
cf.~the claim by Sochor in \cite[p.~142]{Sochor:ASTmeta-III}. 
Our formalization of this fact, presented in the theory \texttt{Not\_regular\_model.thy}, is obtained as an instance of the Bernays-Rieger permutation model given by swapping two elements, namely the empty set $\emptyset$ and its successor $\{\emptyset\}$ (see Section \ref{s:perm} below). 
Under this permutation, the defined membership relation sees $\emptyset$ as a (unique) element of itself, while $\{\emptyset\}$ is the new empty set. All other membership relations are preserved. 
If the universe with the native membership satisfies {\tt (setind)} and {\tt (ts)}, then so does the universe with the defined membership relation.

\smallbreak\noindent
{\bf 2. Axiom of transitive superset} {\tt (ts)}, labelled A82 in \cite{Sochor:ASTmeta-I}, is independent of the fragment axiomatized by
extensionality, empty set, union, powerset, replacement, negated infinity  {\tt (fin)}, and regularity;
that is, by $\zffin$ in which the regularity schema \texttt{(epsind)} has been replaced by ordinary regularity (or \zf{} with \texttt{(fin)} instead of \texttt{(inf)}).
A model showcasing this has been presented in \cite{Mancini-Zambella:noteRecursiveModels}.
The model uses a permutation (on $V_\omega$) analogous to the one presented already in \cite[p.~349]{Rieger:contributionGodelI};
(cf.~also \cite{Enayat-Schmerl-Visser:omegamodels2011}).
The model satisfies all axioms of $\zffin$ with the exception of {\tt (epsind)}, satisfies {\tt (reg)},
and fails {\tt (ts)}. The formalization of these facts is presented in the theory {\tt Not\_ts\_model.thy}.

In a context with {\tt (setext)} and {\tt (sep)}, the axiom ({\tt ts}) yields the existence of a transitive closure of any set.
Further on the positive side, the combination of {\tt (reg)} and {\tt (ts)}, 
in a context of {\tt (setext)} and {\tt (sep)}, 
 proves the axiom schema of regularity ${\tt (regsch)}$.
 Again, this proof can be found in \cite{Sochor:ASTmeta-I} and it proceeds as follows:
let $\varphi$ be an arbitrary formula, and assume that $\varphi(x)$ holds. Let $w$ be the subset of the transitive superset $y$ of $x$ consisting of elements satisfying $\varphi$. If $w$ is empty, then $y$ itself satisfies the conclusion of of \texttt{(regsch)}. If $w$ is not empty, then we have $v \in w$ s.t. $ (v \cap w =  \emptyset)$ by {\tt (reg)}. 
Notice $\varphi(v)$ since $v\in w$. 
By transitivity of $y$, we have $v\subseteq y$, but $v\cap w=\emptyset$,
therefore all elements of $v$ are in $y\setminus w$.
Thus no element of $v$ satisfies $\varphi$.
 In the formalization, the obtained implication is reflected  by the 
 {\tt sublocale\ L\_regsch} command 
 in the locale {\tt  L\_setext\_sep\_reg\_ts}.

\smallbreak\noindent
{\bf 3. Axiom schema of $\in$-induction} {\tt (epsind)}. 
For a given set formula $\varphi$, the instance of {\tt (epsind)}  is logically equivalent 
to the instance of {\tt (regsch)} for $\neg\varphi$. 
The two schemata are therefore interchangeable in no matter which, possibly void, axiomatic context.
In other words,  their equivalence does not depend on the semantics of the symbol $\in$, 
and their equivalence can be established as a pattern in which $\in$ does not occur.  
This is formalized by the lemma {\tt abstract\_foundation\_iff}, established before introducing the set-theoretic signature.

\smallbreak\noindent
{\bf 4. Axiom schema of regularity} \texttt{(regsch)}.
The schema {\tt (regsch)}, labelled  A8 in \cite[p.~709]{Sochor:ASTmeta-I},
entails  {\tt (reg)} just by instantiating the formula in the schema, 
and in the context of
{\tt (setext)}, {\tt (empty)}, {\tt (union)}, {\tt (repl)}, and {\tt (pair)} 
(i.e., in the context of the theory EST of \cite{Baratella-Ferro:theoryNegInf}) it entails  {\tt (ts)}.
This is formalized by the {\tt sublocale L\_ts} expression
in {\tt  L\_setext\_empty\_union\_repl\_pair\_regsch}. 

\smallbreak
 
Summing up, in a sufficiently strong fragment of $\zffin$ without  {\tt (epsind)},
the schema {\tt (epsind)}---or equivalently, {\tt (regsch)}---is equivalent to the combination of {\tt (reg)} and {\tt (ts)}. 
This is formalized in  \texttt{ZFfin.thy}, based on the presentation in \cite{Sochor:ASTmeta-I,Sochor:ASTmeta-II}.
Similarly, the paper \cite{Kaye-Wong:IntpArithmeticSetTheory}) presents equivalence of {\tt (epsind)} and {\tt (ts)}
over the theory EST plus {\tt (reg)}:
but notice that one really need not  assume {\tt (reg)} to obtain {\tt (ts)} from {\tt (epsind)}, since {\tt (epsind)} entails {\tt (reg)} (see the example in Section \ref{s:intro}). Furthermore, \cite[p.~8]{Enayat-Schmerl-Visser:omegamodels2011} misstates the result from \cite{Kaye-Wong:IntpArithmeticSetTheory}, in saying that epsilon induction is equivalent to the existence of transitive closures over EST; but one really needs regularity to establish epsilon induction using the existence of transitive closures in EST (cf.~the independence result mentioned above in {\bf 1.}).

\subsection{Induction and regularity}

Occasionally readers of Vopěnka's work remark casually that they have observed that ``regularity is provable from induction''. 
The axiomatization of the set fragment of the $\atm$ in the paper \cite{Pudlak-Sochor:ModelsAST} amounts to the same effect.
On the other hand, Sochor \cite{Sochor:ASTmeta-III} claims that \texttt{(reg)} is independent of the other axioms,
see also the item {\bf 1.} in the previous subsection. A key to this quandary is the exact formulation of the axiom schema of induction.
The modification of the axiom schema of induction for set formulas for which the above observation is right is labelled
\texttt{(setindregsch)} in our formalization. We show that in  a context identical to the one in which we usually consider 
\texttt{(setind)}, that is,  {\tt (setext)}, {\tt (empty)}, and {\tt (setsuc)} (see subsection \ref{ss:neginf}),
the schema \texttt{(setindregsch)} yields both \texttt{(setind)} and \texttt{(epsind)} (equivalently \texttt{(regsch)}). The former is immediate---\texttt{(setindregsch)} is a stronger variant of \texttt{(setind)}---the latter requires a proof, which is 
formalized with the {\tt sublocale\ L\_setind} commanda
in the locale \texttt{L\_setext\_empty\_setsuc\_setindregsch}.
On the other hand, the axiom schema {\tt (setindregsch)} is provable 
from \texttt{(setext)}, \texttt{(setsuc)}, \texttt{(setind)} and \texttt{(epsind)}. It therefore also holds in the  usual axiomatization of the set fragment of the $\atm$ as given by the locale {\tt ASTset} in the ‹Summary of dependencies› of \texttt{ZFfin.thy}. In that section, we in particular show that the equivalence of our canonical axiomatizations of the theory of hereditarily finite sets, given by locales \texttt{ZFfin} and \texttt{ASTset}, follows directly from previously proven dependencies. The fact can again be verified using Figure \ref{fig:graph-fin}, Figure \ref{fig:graph-reg} and Figure \ref{fig:graph-basic}.  


\subsection{Permutation models}\label{s:perm}

The permutation models method, mentioned above in connection with independence (or ``negative'') results, is based on the following, very general claim. Let $\in$ be a membership relation on a type \texttt{'a } satisfying axioms \texttt{(setext)}, \texttt{(empty)},
\texttt{(power)}, \texttt{(union)}, \texttt{(repl)}, and let $p: \texttt{'a => 'a}$ be a set-definable bijection. Define a new membership relation $\in^p$ by $x \in^p y \equiv x \in p(y)$. Then also $\in^p$ satisfies \texttt{(setext)}, \texttt{(empty)},
\texttt{(power)}, \texttt{(union)}, \texttt{(repl)}. This fact is mentioned and used in \cite{Baratella-Ferro:theoryNegInf} and \cite{Mancini-Zambella:noteRecursiveModels} where it is referred to as ``Fraenkel-Mostowski permutation model''. However, as pointed out in \cite{Enayat-Schmerl-Visser:omegamodels2011}, it can be traced back to Bernays, and was systematically employed by Rieger \cite{Rieger:contributionGodelI}, hence we call it Rieger--Bernays permutation model method. The key fact, which makes the method useful, is that the permutation does not, in general, preserve regularity.

\begin{figure}
	\centering
\begin{tikzpicture}
	\node[compoundN] (notts) at (2,0) {L\_setext\_empty\_power\_repl\_reg + L\_union + L\_setind};
	\node[atomicN] (Lregsch) at (1.5,0.6) {L\_regsch};
	\node[atomicN] (Lts) at (2.5,0.6) {L\_ts};
	\node[compoundN] (notreg) at (2,0.9) {L\_setext\_empty\_power\_repl  + L\_union + L\_setind + L\_ts};
	\node[atomicN] (Lreg) at (2,1.5) {L\_reg};
	\node[compoundN] (notsep) at (5,0.6) {L\_setext + L\_repl};
	\node[atomicN] (Lsep) at (5,1.2) {L\_sep};
	\draw[indE] (Lregsch) -- node[indN] {$\bigotimes$} (notts);
	\draw[indE] (Lts) -- node[indN] {$\bigotimes$} (notts);
	\draw[indE] (Lreg) -- node[indN] {$\bigotimes$} (notreg);
	\draw[indE] (Lsep) -- node[indN] {$\bigotimes$} (notsep);
\end{tikzpicture}
	\caption{Proven facts about independence.
	}
	\label{fig:graph-independence}
\end{figure}
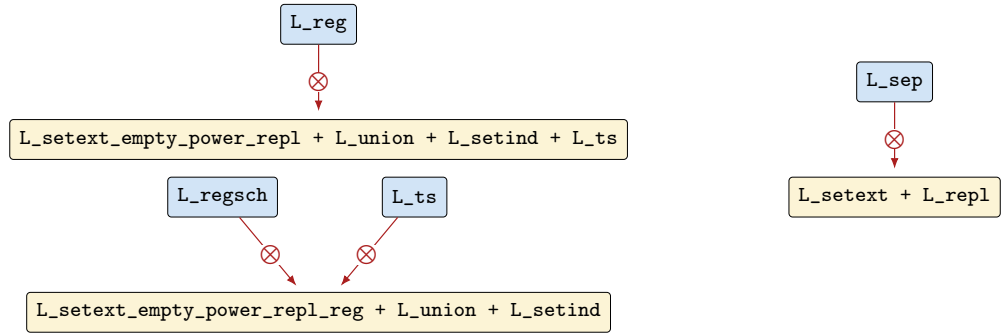

The abovementioned claim is formalized in the theory \texttt{Permutation\_models.thy}. We then use two particular permutations;
see   Figure \ref{fig:graph-independence} for the independence results obtained this way.
The first, simple permutation (introduced already in the previous section), denote it $s$, just swaps $\emptyset$ and $\{\emptyset\}$. 
We also formalize the fact that if $\in$ satisfies \texttt{(ts)} (\texttt{(setind)} resp.) in addition to the above mentioned list of axioms, then also $\in^s$ satisfies \texttt{(ts)} (\texttt{(setind)} resp.). Note that $\emptyset \in^s \emptyset$, therefore the theory with $\in^s$ is not regular. Altogether, we can conclude that the collection of axioms \texttt{(setext)}, \texttt{(empty)},
\texttt{(power)}, \texttt{(union)}, \texttt{(repl)}, \texttt{(ts)} and \texttt{(setind)} is not sufficient to prove \texttt{(reg)} (see theorem \texttt{not\_reg\_model} in \texttt{Not\_regular\_model.thy}).

The second, more complicated permutation $f$, transposing $n$ and the singleton $\{n+1\}$ for all natural numbers $n \ne 0$, is explored in \texttt{Not\_ts\_model.thy}. That permutation is used already by Rieger \cite{Rieger:contributionGodelI}, and again by Mancini and Zambella \cite{Mancini-Zambella:noteRecursiveModels} who stress the fact that the resulting model is recursive.  The construction provides a non-wellfounded membership relation as we have $ \cdots 2 \in^f 1 \in^f 0$. 
Despite of that, the model is regular if the original membership is assumed to satisfy \texttt{(fin)} and \texttt{(reg)}. Then any set contains a finite segment of the infinite chain only, which saves regularity. On the other hand, for the same reason, \texttt{(ts)} is not satisfied, since $\emptyset$ has no transitive superset as it would be infinite if it existed.
 It is a trivial consequence that the model does not satisfy \texttt{(regsch)} (since \texttt{(regsch)} implies \texttt{(ts)}).  Moreover, the model with $\in^f$ can be shown to satisfy $\texttt{(fin)}$.  The final theorem looks like this:   
 
\begin{Icode}
 theorem not_reg_fin_implies_regsch_ts: 
 assumes "L_setext_empty_power_repl_reg (m :: 'a \<Rightarrow> 'a \<Rightarrow> bool)" 
   and "L_fin m" 
 shows "\<not> (\<forall> (mem :: 'a \<Rightarrow> 'a \<Rightarrow> bool). 
  L_setext_empty_power_union_repl_reg_fin mem  \<longrightarrow> 
  L_regsch mem \<or> L_ts mem)"
\end{Icode}
The reader may ask what the role of the assumption is. The answer is related to the fact that Isabelle's simple type theory does not allow to quantify over types. We are therefore unable to say that the implication does not hold for all types (and therefore cannot hold logically, unless the theory is inconsistent). Of course, it would be enough to show a single type on which the implication does not hold. Our claim, however, is stronger, despite the simple type theory: it implies (without quantifying over types explicitly) that the implication does not hold on any type on which a relation can be defined satisfying \texttt{L\_setext\_empty\_power\_repl\_reg\_fin}. Not all types satisfy this condition: namely finite types do not. 

These considerations point out that the membership relation, fixed by~\texttt{set\_signature}, fixes a particular type \texttt{'a} too. And since, as we have just seen, a particular type may have specific properties, one may start to worry about the status of \emph{positive} results. How can the fact that something holds on a particular type imply that it holds logically? The answer to this test of our semantic approach should be clear: the type we talk about is indeed fixed, but at the same time \emph{arbitrary}. We again encounter an implicit quantification over types, facilitated by the type variable. 

\section{Design principles of the formalization}\label{s:formal}

The formalization mirrors fragments of \THF{} by introducing a hierarchy of locales, each given by a particular collection
of axioms or axiom schemata out of a finite list (where an axiom schema counts as one item on the list).
As usual, dependencies between locales are either declared in the definition of a new locale, or proved using the \texttt{interpretation} or \texttt{sublocale} command.

The weakest locale, which in our case bears the name {\tt set\_signature}, introduces only the parameter \texttt{membership} of type \texttt{'a => 'a => bool}, thereby fixing the set-theoretic language as relations on a type {\tt 'a}. The basic language of set theory coincides with the basic language of graph theory, so the formalization can be viewed as
introducing an arbitrary graph with vertices of type {\tt 'a}. 
The strongest theory we consider,  $\zffin$, is captured by two locales, namely \texttt{ZFfin} and \texttt{ASTset},
the equivalence of which is a part of the formalization. Intermediate locales are given descriptive names. For example:
\begin{Icode}
 locale L_union = set_signature + 
  assumes union: "\<forall> x. \<exists> y.\<forall> v. v \<epsilon> y  \<longleftrightarrow>  (\<exists> u. u \<epsilon> x \<and> v \<epsilon> u)"	
\end{Icode}
and 
\begin{Icode}
 locale L_setext_empty_power_union_repl_reg  = 
  L_setext + L_empty + L_power + L_union + L_repl + L_reg
\end{Icode}

The membership relation $\varepsilon$ is intended as the only non-logical symbol of the theory, which means that all other symbols should be expressible in terms of $\varepsilon$ and $=$. This is not literally the case in our formalization due to the fact that we introduce all function symbols already in the locale \texttt{set\_signature}, using Hilbert's \texttt{THE} operator. For example, the function symbol $\emptyset$ is defined by
\begin{Icode}
 definition empty_setM ("\<emptyset>") where "empty_setM \<equiv> collectM (\<lambda> x. x \<noteq> x)"
\end{Icode}
where
\begin{Icode}
 definition collectM :: "('a \<Rightarrow> bool) \<Rightarrow> 'a"  where
  "collectM Q \<equiv> THE z . (\<forall> u . (u \<epsilon> z  \<longleftrightarrow>  Q u))"
\end{Icode}
In other words, the empty set is defined as \emph{the unique} set that contains (``collects'') all sets $x$ with $x \neq x$. 
Properties of Hilbert's \texttt{THE} imply that $\emptyset$ behaves as a fresh constant, unless $\exists! x.\, u \in x \leftrightarrow u \neq u$ is provable.  
This means that while the intended meaning of the definition of $\emptyset$ is clear, we cannot prove $\forall u. \neg(u \in \emptyset)$ outside \texttt{L\_setext\_empty}. Note that the axiom \texttt{(empty)} alone does not suffice either, since without extensionality there may be more than one empty set.
 It would be straightforward to postpone definitions of new notions to contexts where their defining formulas are provable, and the new symbols can be eliminated. Instead we opt for the convenience of defining all symbols at the outset. This way we avoid the need to define the same constant several times in different contexts. 
Moreover, in a few specific situations we obtain claims formulated in the extended language. For example, we have an (easy) claim that \texttt{L\_setind} is a sublocale of \texttt{L\_setindregsch}. This means that \texttt{(setind)} follows from \texttt{(setindregsch)} with no context, in particular, before $\emptyset$ and \texttt{setsucM} have the intended meaning. Consequently, we have a claim that holds in an extended language, and in any structure with arbitrary element and  arbitrary binary function interpreting $\emptyset$ and \texttt{setsucM}, respectively.
In appropriate contexts, we than have a variety of definitional lemmas like
\begin{Icode}
 lemma singleton_def'[set_defs]: "u \<epsilon> {y}\<^sub>M  \<longleftrightarrow>  u = y"
\end{Icode}
Once this is established, the notation $\{y\}_M$ for singleton is guaranteed to have the intended meaning, also in any stronger locales in the hierarchy. We collect such lemmas as named theorems \texttt{set\_defs} so that we can unfold them all at once.

The question of how to deal with the meta-quantification over formulas in schemata of axioms is more tricky. 
Given our semantic approach, we represent formulas of the set language by definable predicates. 
A set formula $\varphi$ with free variables $x_0,x_1,\ldots,x_{n-1}$ defines an $n$-ary predicate  $P_\varphi$, the arguments of which are the values that $x_i$'s in $\varphi$ can take. To accommodate arbitrary arities simultaneously, we work with predicates defined on countably many arguments. 

A cardinal task is to isolate those predicates that correspond to set formulas, excluding predicates that are not first-order definable
in the language of sets, not least in view of our intention to eventually extend our formalization to $\atm$. 
This is crucial for the central aim of the formalization to faithfully represent basic relationships between fragments of the theory $\zffin$. 
We complete the task by defining a predicate \texttt{SetFormulaPredicate} of type  
$\texttt{((nat => 'a) => bool) => bool}$.
That is, we take a predicate $P$ that assigns truth values to maps $\Xi$ of type \texttt{nat => 'a}, i.e., to assignments to countably many variables; and the predicate \texttt{SetFormulaPredicate} applied to any such $P$ indicates whether $P$ corresponds to a set formula.

The predicate \texttt{SetFormulaPredicate} is defined inductively in steps that mimic the construction of a first-order formula as follows:
\begin{Icode}
inductive SetFormulaPredicate :: "((nat \<Rightarrow> 'a) \<Rightarrow> bool) \<Rightarrow> bool" where 
  SFP_mem: "\<And> m n. SetFormulaPredicate (\<lambda> \<Xi>. (\<Xi> m) \<epsilon> (\<Xi> n))"
| SFP_eq: "\<And> m n. SetFormulaPredicate (\<lambda> \<Xi>. (\<Xi> m) = (\<Xi> n))" 
| SFP_neg: "SetFormulaPredicate P \<Longrightarrow> SetFormulaPredicate (\<lambda> \<Xi>. \<not> P \<Xi>)" 
| SFP_disj: "SetFormulaPredicate P \<Longrightarrow> SetFormulaPredicate Q 
    \<Longrightarrow> SetFormulaPredicate (\<lambda> \<Xi>. P \<Xi> \<or> Q \<Xi>)"
| SFP_all: "\<And> n. SetFormulaPredicate P \<Longrightarrow>
    SetFormulaPredicate (\<lambda> \<Xi>. \<forall> a. P (\<Xi>(n:=a)))"
\end{Icode}
Let $P$ be a predicate for which \texttt{SetFormulaPredicate} holds. 
This means that there is a successful run of the inductive definition for $P$, 
and consequently there is a set formula $\varphi_P$ that defines $P$; the latter can be built by tailing the successful run of $\texttt{SetFormulaPredicate}$. 
Consider, for example, the predicate $P_F = \lambda \Xi.\, \texttt{False}$. 
Since \texttt{False} is not directly a set expression, it is not obvious 
that $P_F$ is a set-formula predicate. Nevertheless this is the case, given that 
$$
\lambda \Xi.\, \texttt{False} = \lambda \Xi.\, \Xi\, \texttt{0} \neq \Xi\, \texttt{0} 
$$
is provable, and an application of  \texttt{SFP\_eq} and \texttt{SFP\_neg} gives a successful run of \texttt{SetFormulaPredicate} for the predicate on the right-hand side.
A set-theoretically definable predicate can be defined in multiple ways, which in turn yields multiple distinct runs of \texttt{SetFormulaPredicate}. For example, the predicate 
$P_F$ is induced not only by $x_0 \neq x_0$, but in many other ways, including $x_1 \in x_2 \wedge x_1 \notin x_2$. The latter formula by its structure induces the predicate
$
Q = \lambda \Xi.\, \Xi(1) \in \Xi(2) \wedge  \neg \Xi(1) \in \Xi(2),
$
which turns out to be equal to $P_F$. To show \SFP{} $Q$, we can either prove the equality $Q = P_F$, and reuse the previously proven fact \SFP{} $P_F$, or we can use the expression of $Q$ directly, using the definition of \SFP{}. However, the definition does not include a rule for conjunction. We can therefore either prove an auxiliary rule for conjunction, or we can rewrite $Q$ as 
$
\lambda \Xi.\, \neg (\neg \Xi(1) \in \Xi(2) \vee \Xi(1) \in \Xi(2))
$
and apply the sequence of rules \texttt{SFP\_neg}, \texttt{SFP\_disj}, \texttt{SFP\_neg}, \texttt{SFP\_mem} and \texttt{SFP\_mem}.
In our formalization, we combine these options and again collect useful rules under common names. As a result, we obtain a smooth way of proving that a given predicate is set-formula definable, which typically looks like this:

\begin{Icode}
 have sfp: "SetFormulaPredicate (\<lambda> \<Xi>. \<Xi> 0 = \<emptyset> \<and> \<Xi> 1 = \<Xi> 2)" 
   unfolding logsimps set_defs by (rule SFP_rules)+
\end{Icode}
Here, we want to prove that the predicate ``$x_0$ is empty and $x_1$ is equal to $x_2$'' is set-formula definable. Informally, this seems obvious from the formula $x_0 = \emptyset\, \wedge\, x_1 = x_2$, although even here we have to be careful about the definition of the empty set as discussed above. In any case, proving the fact formally using the definition of \texttt{SetFormulaPredicate} directly would be tedious. Our automation allows us to get close to the informal obviousness.
We use two collections of facts named \texttt{logsimps} (rewriting logical symbols to $\neg$, $\forall$ and $\vee$) and \texttt{set\_defs} (rewriting defined symbols to $\varepsilon$ and $=$). Unfolding those facts transforms the goal into
\begin{Icode}
 SetFormulaPredicate (\<lambda>\<Xi>. \<not> (\<not> (\<forall>u. \<not> u \<epsilon> \<Xi> 0) \<or> \<Xi> 1 \<noteq> \<Xi> 2))
\end{Icode}
which is then discharged by repeated application of the third collection of named facts \texttt{SFP\_rules}.

Provided that $P$ is a \texttt{SetFormulaPredicate}, it can be proven by induction on the structure of $P$ (see the lemma \texttt{bounded\_free}) that the truth value of $P$ on  $\Xi$ depends on a finite number of values only, corresponding to free variables which must appear in any formula that represents the predicate $P$.
For convenience and readability, we also introduce \texttt{SetProperty} and \texttt{SetRelation} for unary and binary set-formula predicates.

 With this setup, we can formulate axiom schemata as desired. For example, the schema of induction looks like this:
\begin{Icode}
 locale L_setind = set_signature +
  assumes setind: "\<And> P. SetFormulaPredicate P  \<Longrightarrow> 
   \<forall> x. P (\<Xi>(0:= \<emptyset>))  \<longrightarrow>  
     (\<forall> x y. P (\<Xi>(0:= x))  \<longrightarrow>  P (\<Xi>(0:= setsucM x y)))  \<longrightarrow>  P (\<Xi>(0:= x))"
\end{Icode}
The variable over which the induction is performed is $x_0$ of type \texttt{'a}. The assignment $\Xi(0 := x)$ corresponds to the substitution of the value $x$ for $x_0$. Note that $\Xi$ is free in the axiom, which corresponds to the fact that a formula $\varphi$ in the classical formulation of the induction schema can contain free variables (an aspect which often remains tacit).

In order to further illustrate the use of \texttt{SetFormulaPredicate}, we look at lemma \texttt{setind\_var} (in the theory \texttt{ZFfin.thy}), which allows to perform the induction on an arbitrary variable $x_n$, a possibility that goes without saying in a paper proof. That is, the lemma says:

\begin{Icode}
 lemma setind_var:   
  assumes "SetFormulaPredicate P" and "P(\<Xi>(n:= \<emptyset>))" and  
     step: "\<forall> x y. P(\<Xi>(n:= x))  \<longrightarrow>  P (\<Xi>(n:= setsucM x y))"
  shows "P(\<Xi>(n:= x))"
\end{Icode}
The task is to rename variables so that the locale axiom \texttt{setind} can be used.
We first obtain the index $m$ such that the value of $P$ does not depend on any variable $x_j$, $m\le j$,
\begin{Icode}
 from bounded_free[OF \<open>SetFormulaPredicate P\<close>]
 obtain m where m_def: "P \<Xi> = P \<Xi>'" if "\<forall>i<m. \<Xi> i = \<Xi>' i" for \<Xi> \<Xi>'
   by blast
\end{Icode}
This means that there is a set-formula $\varphi(x_0,x_1,\ldots,x_{m-1})$ which defines the value $P(\Xi)$ by substituting $\Xi(i)$ for $x_i$.
We then define an assignment $X$ which modifies $\Xi$ by duplicating the value $\Xi(0)$ at the position $m+n+1$, which is a position both corresponding to a fresh variable of $\varphi$, and distinct from $n$.  
\begin{Icode}
  let ?X = "\<Xi>(Suc (n + m) := \<Xi> 0)"
\end{Icode}
Let $f$ be the mapping  $f: 0 \mapsto (n+m+1)$, $n \mapsto 0$, and identity otherwise.
\begin{Icode}
  let ?f = "id(0 := Suc (n + m), n := 0)" 
\end{Icode}
Consider a new predicate $Q$
\begin{Icode}
  let ?Q = "\<lambda> \<Xi>. (P (\<lambda> b. \<Xi> (?f b)))"
\end{Icode}
such that the value of $Q$ on the assignment $\Xi$ is given by 
substituting $\Xi(f(b))$ for $x_b$ in $\varphi$. In other words, the predicate $Q$ is defined by the formula $\varphi(x_{f(0)}, x_{f(1)},\dots,x_{f(m-1)})$. Formalizing this insight, we obtain 
 lemma \texttt{transform\_variables} claiming that an arbitrary transformation of indices in a set-definable predicate yields again a set-definable predicate. Using that lemma we obtain
\begin{Icode}
 have sfpq: "SetFormulaPredicate ?Q"
  using transform_variables[OF \<open>SetFormulaPredicate P\<close>] by simp	
\end{Icode}
We also have the following claim
\begin{Icode}
 have small: "\<forall> i<m. (?X (0 := u)) (?f i) = (\<Xi>(n := u)) i" for u
  by auto
\end{Icode}
That is, $X'(f(i)) = \Xi'(i)$, $i < m$, where $\Xi'$ modifies $\Xi$ by assigning (an arbitrary) set $u$  to the index $n$, and $X'$ modifies $X$ by assigning (the same) $u$ to $0$. The verification is straightforward using definitions. It is trivial for $i$ distinct from $0$ and $n$, since then no updates apply, $f(i) = i$ and $X(i) = \Xi(i)$.
For $i = 0$ or $i = n$, assume first $n \neq 0$. Then we have
 $X'(f(0)) = X'(m+n+1) = X(m+n+1) = \Xi(0) = \Xi'(0)$ and $X'(f(n)) = X'(0) = u = \Xi'(n)$. If $n = 0$, then $f$ is identity everywhere (the value of $f(0)$ is set  to $n+m+1$, and then restored back to $0$) and we have $X'(f(i)) = u = \Xi'(i)$ for $i = 0 = n$. Note however, that for $n = 0$, there is nothing to prove since \texttt{setind\_var} is just \texttt{setind}. 
In a paper proof, one would probably start the proof by assuming (perhaps tacitly) that $n\neq 0$. In the formalized proof, we noticed that the assumption is superfluous.

Keeping in mind the range of variables that can influence the value of $P$, it is equally straightforward to verify that $P$ has the same value on $\Xi'$ as $Q$ has on $X'$. In fact, this is just \texttt{small} after using the defining property of $m$ for $\Xi'$ and $X'$.
\begin{Icode}
 have equiv: "(?Q (?X (0 := u)))  \<longleftrightarrow>  (P (\<Xi>(n := u)))" for u
  by (rule m_def[of "\<Xi>(n := u)" "\<lambda> b. (?X (0 := u))(?f b)"]) 
     (rule small) 
\end{Icode}
The claim of \texttt{setind\_var}, which is $P(X')$ for $u = x$, now follows by first replacing $P(\Xi')$ with $Q(X')$, and then using the axiom \texttt{setind} with $Q$ and $X$:
\begin{Icode}
  show  "P(\<Xi>(n:= x))"
   unfolding equiv[symmetric]
   by (rule setind[rule_format, OF sfpq, of ?X], unfold equiv)
   (use assms in blast)+
\end{Icode}

\section{Related work}
\label{s:related}

Two distinct approaches to $\zf$ are available in Isabelle, given its generic nature. The first one is Isa\-belle/ZF, formalizing results directly in $\zf$. This covers many advanced results, including forcing \cite{Forcing-AFP}. 
Another approach is dealing with $\zf$ within HOL. The latter approach itself has two instances. One is Obua's $\zf$ library in the main distribution of Isabelle/HOL, the second one is Paulson's AFP entry \cite{ZFCinHOL-AFP}, which identifies classes with the type \texttt{V set} and singles out sets as elements corresponding to small classes. It is not clear how any of these developments could be used for our purposes.

More generally, the two approaches exist with respect to FOL. Apart from Isabelle/FOL, there are several developments using HOL in order to obtain results about FOL (and other logics). A prominent example is Paulson's formalization of G\"odel incompleteness theorems \cite{Incompleteness-AFP,Incompleteness-jar}. The latter development is based on the formalization \texttt{HereditarilyFinite} \cite{HereditarilyFinite-AFP}, 
obtained by defining the membership relation \texttt{hmem} on the abstract type \texttt{hf} based on \texttt{nat} using 
the main idea behind the celebrated Ackermann interpretation, eventually yielding one direction of the bi-interpretability result between $\pa$ and $\zffin$.  In our formalization we verify that the membership relation \texttt{hmem} satisfies axioms of $\zffin$: 
\begin{Icode}
 interpretation zfhf: ZFfin "(\<boldin>)"
  rewrites "zfhf.emptysetM = 0" and
     "zfhf.singletonM y = \<lbrace>y\<rbrace>" and
     "zfhf.setsucM x y = x \<triangleleft> y"
\end{Icode}
Paulson's formalization of incompleteness theorems uses nominal types in order to reason about the syntax of first-order theory. 
This is a major difference from our approach, which uses locales to capture axioms and can be described as semantic 
(however, Paulson's approach is itself described as semantic in \cite{PopescuTraytel2019}).
On a related note, Paulson's theory HF cannot be scaled down to subtheories, as it is built entirely on 
the said relation on the primitive type \texttt{nat}.

Arguably, because of the bi-interpretability result, theories of prime interest are $\pa$ and its subtheories. On that score,
a formalization of Robinson's arithmetic $\qa$ \cite{Robinson-Arithmetic-AFP} comes as a part of a broader project
of Popescu and Traytel \cite{SyntaxIndependentLogic-AFP,PopescuTraytel2019}, which provides a general framework for 
formalizing different logics using Isabelle/HOL. 
They do not use nominal types but axiomatize syntactic notions (such as free variables) by some general properties. 

We believe that comparing our fully semantic approach with approaches mentioned above is of an independent interest going beyond the scope of this paper.

\section{Concluding remark}
\label{s:final}

The formalization is open to additions (see, for example, the model mentioned in Section \ref{ss:neginf}, \textbf{C}). As already mentioned, the fragments that do not use the finiteness axioms can be extended with the axiom 
of infinity {\tt (inf)}, so one can develop $\zf$ in this way.
Furthermore, one can adapt the principles that we rely on to formalize subsystems of Peano arithmetic.
Our broad intent, beyond completing details of the theory of hereditarily finite sets, 
 would be to extend the formalization to deal with classes, with an outlook of formalizing Vopěnka's $\atm$.  
 
 The authors are grateful to three anonymous reviewers for their comments.

\bibliographystyle{eptcs}
\bibliography{vopenka_merge}

\end{document}